\documentclass[%
 reprint,
 amsmath,amssymb,
 aps,
twocolumn,
nofootinbib
]{revtex4}

\usepackage{graphicx}
\usepackage{dcolumn}
\usepackage{bm}

\usepackage{mathrsfs} 
\newcommand{\scri}{{\mathscr I}}

\begin{document}

\preprint{CSCVR-12-2013}

\title{Brief note on high-multipole Kerr tails}

\author{Tyler Spilhaus}
\author{Gaurav Khanna}%
\email{gkhanna@umassd.edu}
\affiliation{%
Physics Department and Center for Scientific Computing and Visualization Research, \\
University of Massachusetts Dartmouth, North Dartmouth, MA 02747
}%

\begin{abstract}
In this note we reconsider the late-time, power-law decay rate of scalar fields in a Kerr space-time background. We implement a number of mathematical and computational enhancements to our time-domain, (2+1)D Teukolsky evolution code and are able to obtain reliable decay rates for multipoles as high as $\ell=16$. Our numerical results suggest full agreement with the proposed decay expressions in recent work~\cite{ZKB,BK} for both the finite distance (including horizon), and null infinity cases. We also study the same in the context of extremal Kerr space-time and find that the same results hold, except that the horizon tails follow the null infinity expressions instead.
\end{abstract}

\date{\today}

\maketitle


\section{Background \& Summary}
The asymptotic late-time, power-law decay ($t^n$) rates of matter fields in Kerr black hole space-time has been a matter of some debate for several decades. However, significant progress has been made on this question over the past few years through the development of sophisticated techniques~\cite{Tiglio,GPP,AnilZ,Revisited,Rakesh,ZKB,RT,Harms} to tackle the computational challenges inherent to this problem. Recent work~\cite{ZKB,BK} proposes the following late-time decay rate expressions for the scalar field case:  
\begin{equation}  \label{eq:rates}\quad
n = \left\{ \begin{array}{ll}
-(\ell'+\ell + 3) & \mathrm{for}\quad \ell'=0, 1 \\
-(\ell'+\ell + 1) & \mathrm{otherwise}\quad \end{array}\right.\\
\end{equation}
and 
\begin{equation} \label{eq:rates2}\quad
n^{\scri^+} = \left\{ \begin{array}{ll}
-\ell' & \mathrm{for}\quad \ell\leq \ell'-2  \\
-(\ell+2) & \mathrm{for} \quad \ell\geq \ell' \end{array}\right.
\end{equation}
by carefully studying the ``inter-mode coupling'' effects that are present in Kerr space-time due to frame-dragging. Note that these expressions above are for the axisymmetric multipoles. $\ell'$ refers to the initial field multipole and $\ell$ is the multipole of interest under study.

In this note, we borrow and implement the ``best-practice'' lessons from all previous numerical work on Kerr tails (especially~\cite{RT,Rakesh,Khanna}) and perform very high-accuracy computations for the axisymmetric multipoles up to $\ell=\ell'=16$. Our results are in full agreement with the expressions \ref{eq:rates}, \ref{eq:rates2} above. It should be noted that we are not presenting any new physical results in this work or developing a deeper understanding of known expressions; we are simply demonstrating a proposed result in the literature~\cite{ZKB,BK} to be accurate for a large range of parameters.  

\section{Methodology}
We begin this section by briefly commenting on why numerical computations are so challenging in the context of Kerr tails: (i) These simulations are required to be rather long duration -- this is because typically the observed field exhibits an exponentially decaying oscillatory behavior in the initial part of the evolution i.e. quasi-normal ringing, and only much later does this transition over to a clean power-law decay. Therefore, one needs to evolve past the point that the initial oscillations decay away. Moreover, as clearly shown in Refs.~\cite{ZKB,BK} there is an ``intermediate'' tail regime, wherein one observes tails with various decay rates that are not necessarily the late-time asymptotic rates that we are after. Indeed, this intermediate regime is largely responsible for much of the confusion on this topic, in the literature. These intermediate tails decay faster than the asymptotic rate, but typically have dominant amplitudes for a period of time. We must evolve past this regime as well, in order to obtain the true asymptotic rate; (ii) Because each multipole has its own decay rate (which increases with an increase in $\ell$) at late times one obtains numerical data in which different multipoles have widely different amplitudes (often 30 -- 40 orders of magnitude apart!). For this reason, the numerical solution scheme is required to have high-order convergence (to reduce the discretization errors to low enough levels, to be able to track the fast decaying multipoles); (iii) Moreover, these computations also require high-precision floating-point numerics, due to the very large range of amplitudes involved, and also to reduce round-off error which can easily overwhelm the fast decaying modes. 

To address the challenges mentioned above, we perform the following mathematical and computational advancements to our time-domain, (2+1)D Teukolsky equation evolution code:

\subsection{Mathematical Enhancements}
The main advance we make in this context is to recast the problem using the technique of hyperboloidal compactification~\cite{hyper} for the Teukolsky equation in Kerr space-time. This allows one to include null infinity $\scri^+$ on the computational grid by mapping the entire space-time onto a compact domain. This technique also allows us to use a rather modest sized grid to sample the entire domain, thus delivering tremendous savings towards the total computational cost. In short, hyperboloidal compactification enables us to perform long duration simulations using relatively modest computational resources. In this work, we use the compactification approach similar to Ref.~\cite{RT,Harms,hyper2} that utilizes a single horizon-penetrating hyperboloidal foliation with conformal compactification. More specifically, we utilize the compactified radial coordinate $\rho$ given by $r = \rho/\Omega$, and $\Omega = 1-\rho/S$, where $S$ is the location of the outer boundary. For all our numerical simulations, we choose a Kerr hole with $a/M = 0.8$ and use $S/M=38.4$, with the inner boundary at the (outer) horizon.  

\begin{figure}[!t]
\centering
\includegraphics[width=3.5in]{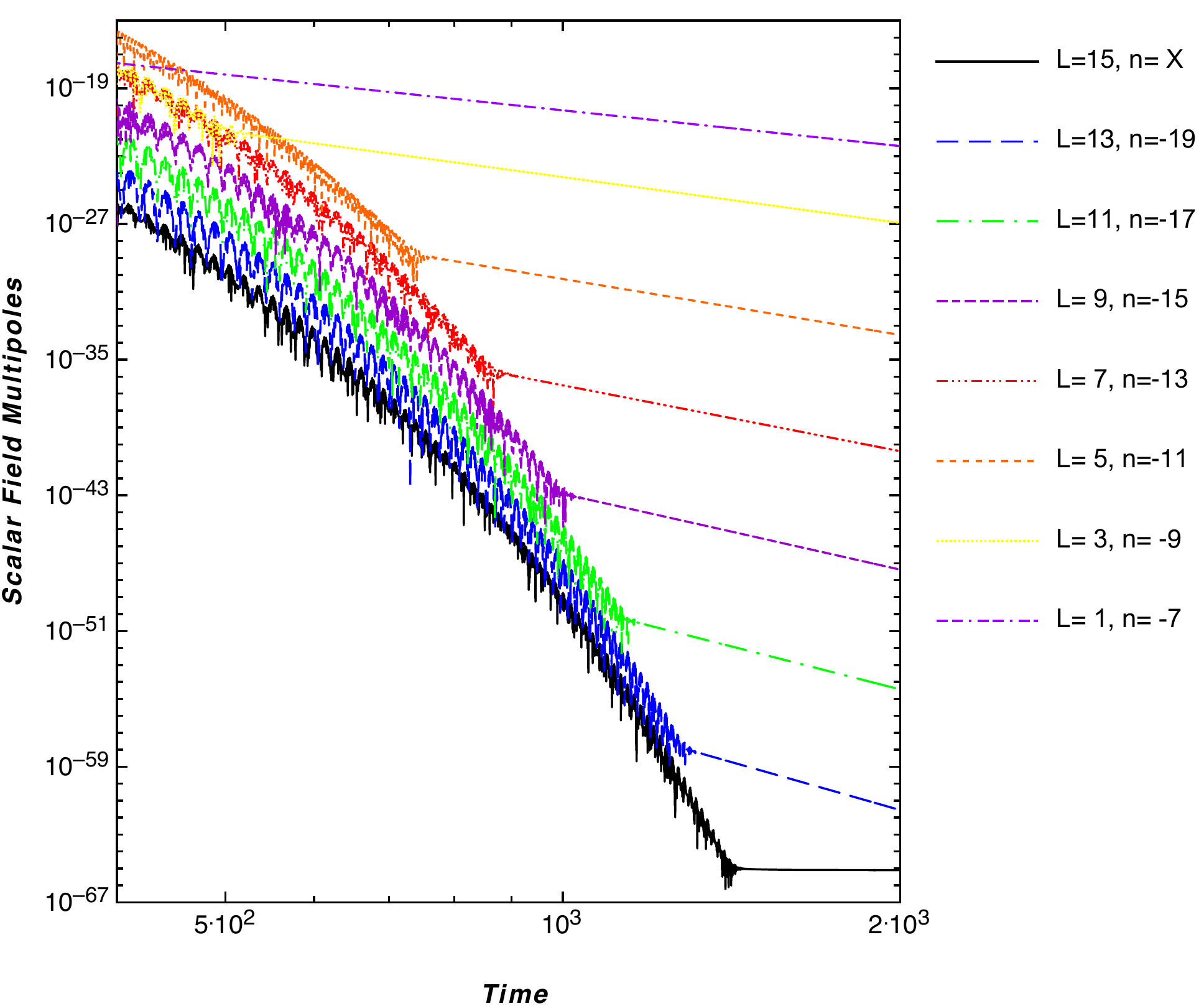}
\caption{Kerr tails for a range of $\ell$ multipoles (1 -- 15) starting with a pure $\ell' = 5$ multipole. These tails obey the proposed tail formula $n=-(\ell + \ell' + 1)$. It is clear that such numerical simulations require octal-precision floating-point arithmetic. } 
\end{figure}

\subsection{Computational Enhancements}
We make three additional major enhancements to our Teukolsky code in order to obtain the high-accuracy numerical results we report in the following section. First, a pseudo-spectral collocation numerical scheme is implemented in order to accurately and efficiently handle the spatial discretization. Because the tail solutions are smooth, this method converges {\em exponentially}, thus allowing us to significantly lower the discretization error as needed, even with a modest grid size. We utilize a single spectral domain in this work, with the angular and radial directions expanded in Legendre and Chebyshev polynomials, respectively, using the Gauss-Lobatto collocation points. We use $245$ collocation points in the radial direction and $16$ in the angular, for all the simulations in this work. We use the method-of-lines approach to evolve forward in time, using a fourth-order Runge-Kutta method. The time-step we use in all our computations is $\Delta t/M = 0.05$. Secondly, we implement high-precision, floating-point arithmetic throughout our code. In particular, this enhancement includes full support for {\em octal-precision numerics} (256-bit or $\sim$60 decimal digits). This is required to keep the round-off error in our simulations at acceptable levels. Finally, we also implement algorithmic parallelism in order to speed up the computations, so they complete in a reasonable amount of time. In particular, we use a small cluster of 16 Sony PlayStation 3 gaming consoles (PS3)~\footnote{http://gravity.phy.umassd.edu/ps3.html} to perform all the simulations in this work. Each PS3 works independently on an initial multipole $\ell'$ case (an ``embarrassing'' coarse-grain parallelism). In addition, we also implement a fine-grain parallelism at the level of the high-precision computations themselves (see Ref.~\cite{Rakesh,Khanna} for details) utilizing the parallel architecture of the PS3's Cell Broadband Engine. Overall, we obtain over {\em two orders-of-magnitude speed-up} via this parallel computing approach. 

In summary, pseudo-spectral collocation method enables us to drastically reduce the discretization error, high-precision numerics helps us accurately track amplitudes down to the $10^{-60}$ scale, and the parallelism and compactification help to keep the total runtime to stay reasonable (and also obtain rates at null infinity directly). 

\begin{table}[h]
  \begin{center}
      \begin{tabular}{|c||c|c|c|c|c|c|c|c|c|}
        \hline
        $\ell' \backslash \ell$ & 0 & 2 & 4 & 6 & 8 & 10 & 12 & 14 & 16 \\
        \hline\hline 
        0                       &-3 &-5 &-7 &-9 &-11&-13 &-15 &-17 & X   \\
        2                       &-3 &-5 &-7 &-9 &-11&-13 &-15 &-17 & X   \\
        4                       &-5 &-7 &-9 &-11&-13&-15 &-17 &-19 & X   \\
        6                       &-7 &-9 &-11&-13&-15&-17 &-19 &  X & X   \\
        8                       &-9 &-11&-13&-15&-17&-19 &  X &  X & X   \\
        10                      &-11&-13&-15&-17&-19&  X &  X &  X & X   \\
        12                      &-13&-15&-17&-19&-21&-23 &  X &  X & X   \\
        14                      &-15&-17&-19&-21&-23&-25 &  X &  X & X   \\
        16                      &-17&-19&-21&-23&-25&  X &  X &  X & X   \\
        \hline
   \end{tabular}
  \end{center}
\caption{Asymptotic late-time scalar field tails in Kerr space-time at finite distances for even multipoles.}
\end{table}

\begin{table}[h]
  \begin{center}
      \begin{tabular}{|c||c|c|c|c|c|c|c|c|c|}
        \hline
        $\ell' \backslash \ell$ & 1 & 3 & 5 & 7 & 9 & 11 & 13 & 15 \\
        \hline\hline 
        1                       &-5 &-7 &-9 &-11&-13&-15 &-17 &-19 \\
        3                       &-5 &-7 &-9 &-11&-13&-15 &-17 &-19 \\
        5                       &-7 &-9 &-11&-13&-15&-17 &-19 &  X \\
        7                       &-9 &-11&-13&-15&-17&-19 &-21 &  X \\
        9                       &-11&-13&-15&-17&-19&  X &  X &  X \\
        11                      &-13&-15&-17&-19&-21&  X &  X &  X \\
        13                      &-15&-17&-19&-21&-23&  X &  X &  X \\
        15                      &-17&-19&-21&-23&-25&-27 &  X &  X \\
        \hline
   \end{tabular}
  \end{center}
\caption{Asymptotic late-time scalar field tails in Kerr space-time at finite distances for odd multipoles.}
\end{table}

\section{Numerical Results}
In this section, we present the outcome of the enhancements implemented in the previous section. 

\subsection{Non-Extremal Kerr}
The initial data is a smooth Gaussian wave-packet centered at $\rho/M=3.0$ and of width $\sigma/M=2.0$ and the time evolution terminates at $t/M=2000$. The results are detailed in four tables: two for the finite-distance rates (odd and even multipole cases separately) and another two tables for the null infinity rates. It is worth pointing out the tails on the horizon follow the same behavior as the ones at finite-distance. Except for a few cases (close to $\ell=16$) where even octal-precision numerics prove to be insufficient to produce reliable tail solutions, all results agree precisely with expressions \ref{eq:rates}, \ref{eq:rates2} above. The tables are self-explanatory. All depicted numerical results are accurate within a few percent or less. Figure 1 depicts the actual simulation data for a sample $\ell'=5$ case.  

\begin{table}[h]
  \begin{center}
      \begin{tabular}{|c||c|c|c|c|c|c|c|c|c|}
        \hline
        $\ell' \backslash \ell$ & 0 & 2 & 4 & 6 & 8 & 10 & 12 & 14 & 16 \\
        \hline\hline 
        0                       &-2 &-4 &-6 &-8 &-10&-12 &-14 &-16 &  X   \\
        2                       &-2 &-4 &-6 &-8 &-10&-12 &-14 &-16 &  X   \\
        4                       &-4 &-4 &-6 &-8 &-10&-12 &-14 &-16 &-18   \\
        6                       &-6 &-6 &-6 &-8 &-10&-12 &-14 &-16 &-18   \\
        8                       &-8 &-8 &-8 &-8 &-10&-12 &-14 &-16 &-18   \\
        10                      &-10&-10&-10&-10&-10&-12 &-14 &-16 &-18   \\
        12                      &-12&-12&-12&-12&-12&-12 &-14 &-16 &-18   \\
        14                      &-14&-14&-14&-14&-14&-14 &-14 &-16 &-18   \\
        16                      &-16&-16&-16&-16&-16&-16 &-16 &-16 &-18   \\
        \hline
   \end{tabular}
  \end{center}
\caption{Asymptotic late-time scalar field tails in Kerr space-time at null infinity for even multipoles.}
\end{table}

\begin{table}[h]
  \begin{center}
      \begin{tabular}{|c||c|c|c|c|c|c|c|c|c|}
        \hline
        $\ell' \backslash \ell$ & 1 & 3 & 5 & 7 & 9 & 11 & 13 & 15 \\
        \hline\hline 
        1                       &-3 &-5 &-7 &-9 &-11&-13 &-15 &-17 \\
        3                       &-3 &-5 &-7 &-9 &-11&-13 &-15 &-17 \\
        5                       &-5 &-5 &-7 &-9 &-11&-13 &-15 &-17 \\
        7                       &-7 &-7 &-7 &-9 &-11&-13 &-15 &-17 \\
        9                       &-9 &-9 &-9 &-9 &-11&-13 &-15 &-17 \\
        11                      &-11&-11&-11&-11&-11&-13 &-15 &-17 \\
        13                      &-13&-13&-13&-13&-13&-13 &-15 &-17 \\
        15                      &-15&-15&-15&-15&-15&-15 &-15 &-17 \\
        \hline
   \end{tabular}
  \end{center}
\caption{Asymptotic late-time scalar field tails in Kerr space-time at null infinity and for odd multipoles.}
\end{table}

It took approximately two days to perform all the computations we have presented in this section, using the 16 PS3 cluster we mentioned before. 

\subsection{Extremal Kerr}
For obtaining reliable results for the extremal case ($a/M = 1$) we use the same settings as in the previous subsection, however with twice as many collocation points. The finite-distance and null infinity rates obtained in this case are identical to the results presented in the last subsection, so we will not repeat those here. However, the horizon rates differ completely and follow the null infinity expressions instead. Due to computational resource limitations we are only able to accurately obtain data for a small subset of the ($\ell'€™$, $\ell$) cases considered thus far. Those results are presented in the two tables below. 

\begin{table}[h]
  \begin{center}
      \begin{tabular}{|c||c|c|c|c|c|c|c|c|c|}
        \hline
        $\ell' \backslash \ell$ & 0 & 2 & 4 & 6 & 8 & 10 & 12 & 14 & 16 \\
        \hline\hline 
        0                       &-2 &-4 &-6 &-8 &-10&-12 &-14 &  X &  X \\
        2                       &-2 &-4 &-6 &-8 &-10&-12 &  X &  X &  X \\
        \hline 
        \hline
        $\ell \backslash \ell'€™$ & 0 & 2 & 4 & 6 & 8 & 10 & 12 & 14 & 16 \\
        \hline\hline 
        0                       &-2 &-2 &-4 &-6 &-8&   X &  X &  X &  X \\
        \hline
   \end{tabular}
  \end{center}
\caption{Asymptotic late-time scalar field tails in extremal Kerr space-time at the horizon for even multipoles.}
\end{table}

\begin{table}[h]
  \begin{center}
      \begin{tabular}{|c||c|c|c|c|c|c|c|c|c|}
        \hline
        $\ell' \backslash \ell$ & 1 & 3 & 5 & 7 & 9 & 11 & 13 & 15 \\
        \hline\hline 
        1                       &-3 &-5 &-7 &-9 &-11&-13 &-15 &  X \\
        \hline 
        \hline
        $\ell \backslash \ell'€™$ & 1 & 3 & 5 & 7 & 9 & 11 & 13 & 15 \\
        \hline\hline 
        1                       &-3 &-3 &-5 &-7 & X &  X &  X &  X \\
        \hline
   \end{tabular}
  \end{center}
\caption{Asymptotic late-time scalar field tails in extremal Kerr space-time at the horizon for odd multipoles.}
\end{table}

The tails at the event horizon matching those at null infinity may seem surprising at first, but, it should be noted that this feature has been appreciated before in the literature, at least for extremal Reissner-Nordstrom black holes~\cite{symm}. This happens as a result of a discrete conformal symmetry between the horizon and null infinity in these space-times. Our numerical results are simply an interesting and independent verification of this symmetry in the context of extremal Kerr space-time.  

\section{Acknowledgements}
We would like to thank Anil Zenginoglu for providing us the details of the hyperboloidal compactification we utilized in this work. We would also like to thank Manuel Tiglio for providing us the pseudo-spectral collocation code used in Ref.~\cite{Tiglio}, which we enhanced and repurposed for this work. We also acknowledge feedback provided by Lior Burko on this note. G.K. acknowledges research support from NSF Grant Nos. PHY-1016906, CNS-0959382, PHY-1135664, PHY-1303724 and PHY-1414440, and from the US Air Force Grant Nos. FA9550-10-1-0354 and 10-RI-CRADA-09. G.K. is also grateful to Snowy for his constant support over a decade. T.S. acknowledges research support from NSF Grant Nos. PHY-1016906 and PHY-1414440, and also the Physics Department of the University of Massachusetts Dartmouth.


\begin{thebibliography}{99}

\bibitem{Tiglio} Manuel Tiglio, Lawrence Kidder, Saul Teukolsky: ``High accuracy simulations of Kerr tails: coordinate dependence and higher multipoles'',  Class. Quant. Grav. 25, 105022 (2008).

\bibitem{GPP} Reinaldo J. Gleiser, Richard H. Price, Jorge Pullin: ``Late time tails in the Kerr spacetime'', Class. Quant. Grav. 25, 072001 (2008).

\bibitem{AnilZ} Anil Zenginoglu, Manuel Tiglio: ``Spacelike matching to null infinity'', Phys. Rev. D 80, 024044 (2009).

\bibitem{Revisited} Lior M. Burko, Gaurav Khanna: ``Late-time Kerr tails revisited'', Class. Quant. Grav. 26, 015014 (2009).

\bibitem{Rakesh} Rakesh Ginjupalli, Gaurav Khanna: ``High-Precision Numerical Simulations of Rotating Black Holes Accelerated by CUDA'', Proceedings of the  International Conference on High Performance Computing Systems (HPCS), Orlando, FL (2010).

\bibitem{ZKB} Anil Zenginoglu, Gaurav Khanna, Lior M. Burko: ``Intermediate behavior of Kerr tails", to appear in Gen. Rel. Grav. (2013).

\bibitem{RT} Istvan Racz, Gabor Z. Toth: ``Numerical investigation of the late-time Kerr tails'', Class. Quant. Grav. 28, 195003 (2011).

\bibitem{Harms} Enno Harms, Sebastiano Bernuzzi, Bernd Bruegmann: ``Numerical solution of the 2+1 Teukolsky equation on a hyperboloidal and horizon penetrating foliation of Kerr and application to late-time decays'', 	Class. Quant. Grav. 30, 115013 (2013).

\bibitem{BK} Lior M. Burko, Gaurav Khanna: ``Mode coupling mechanism for late-time Kerr tails'', submitted to Phys. Rev. D (2013).

\bibitem{Khanna} Gaurav Khanna: ``High-Precision Numerical Simulations on a CUDA GPU: Kerr Black Hole Tails'', J. of Sci. Comp. 56, 366 (2013).

\bibitem{hyper} Anil Zenginoglu: ``Hyperboloidal foliations and scri-fixing'', Class. Quant. Grav. 25, 145002 (2008).

\bibitem{hyper2} Anil Zenginoglu: ``A hyperboloidal study of tail decay rates for scalar and Yang-Mills fields'', Class. Quant. Grav. 25, 175013 (2008).

\bibitem{symm} Piotr Bizon, Helmut Friedrich: ``A remark about wave equations on the extreme Reissner-Nordstrom black hole exterior'', Class. Quant. Grav. 30, 065001 (2013). 

\end{thebibliography}
\end{document}